\documentclass[aps,prb,twocolumn,superscriptaddress,epsfig,amsmath,showpacs]{revtex4-1}

\usepackage{epsfig}
\usepackage{dcolumn}
\usepackage{multirow}
\usepackage{bm}
\usepackage{color}
\usepackage{soul}
\setstcolor{red}
\setulcolor{red}
\begin{document}
\title{Effects of spin-orbit interaction on magnetic and electronic structures in antiferromagnetic LaFeAsO}

\author{Sehoon Oh}
\affiliation{Department of Physics, IPAP, and Center for Computational Studies of Advanced Electronic Material Properties, Yonsei University, Seoul 03722, Korea}
\author{Jung-Hoon Lee}
\affiliation{Department of Materials Science and Engineering, and Division of Advanced Materials Science, Pohang University of Science and Technology, Pohang 37673, Korea}
\author{Hyun Myung Jang}
\affiliation{Department of Materials Science and Engineering, and Division of Advanced Materials Science, Pohang University of Science and Technology, Pohang 37673, Korea}
\affiliation{Department of Physics, Pohang University of Science and Technology, Pohang 37673, Korea}
\author{Hyoung Joon Choi}
\email[Email:\ ]{h.j.choi@yonsei.ac.kr}
\affiliation{Department of Physics, IPAP, and Center for Computational Studies of Advanced Electronic Material Properties, Yonsei University, Seoul 03722, Korea}

\date{\today}

\begin{abstract}
Magnetic and electronic structures in LaFeAsO in the single-stripe-type antiferromagnetic (AFM) phase are studied using first-principles 
density-functional calculations including the spin-orbit interaction. 
We show that the longitudinal ordering (LO) where Fe magnetic moments are parallel or anti-parallel with the in-plane AFM ordering vector is lower in energy than transverse orderings (TOs), in good 
agreement with neutron diffraction experiments.
Calculated energy difference between LO and TOs is about 0.1~meV per Fe atom, indicating that LO will prevail at temperature below about 1~K.
We also show that the spin-orbit interaction splits degenerate bands at some
high-symmetry points in the Brillouin zone by about 60 meV, depending on spatial directions of the Fe magnetic moments.
\end{abstract}


\maketitle

Role of magnetism in the superconductivity of iron oxypnictides LaFeAsO and other \textit{R}FeAsO (\textit{R} = rare-earth) has been a central issue since their discovery.\cite{1, 2, 3}
These materials exhibit superconductivity at high temperatures, for example, 55~K in SmFeAsO$_{1-x}$F$_x$.\cite{4} 
Without doping, these compounds show antiferromagnetism at low temperature, and they become superconductors when doped with electrons or holes.
The emergence of superconductivity in the vicinity of magnetic phase suggests that understanding the magnetic properties may provide a clue to the origin of their unconventional superconductivity.\cite{5,6,yildirim,7,8,9,10,11,12,13,Review_P,Review_O,Review_B,Review_D}

As temperature lowers, undoped LaFeAsO undergoes a structural transition from a tetragonal to an orthorhombic structure and Fe magnetic moments form a single-stripe-type antiferromagnetic (AFM) ordering, where Fe magnetic moments are ordered antiferromagnetically in an orthorhombic in-plane axis, say $a$ axis, ferromagnetically in the other in-plane axis, say $b$ axis, and antiferromagnetically along the $c$ axis.\cite{6}
Density functional theory (DFT) calculations have been successful in predicting and explaining the AFM wave vector in iron pnictides\cite{5,yildirim,7,9,11,cao,14,15} and chalcogenides\cite{16,17} although the DFT calculations predict substantially larger values of Fe magnetic moments than measurements.\cite{6,18}

Neutron diffraction experiments\cite{18} on the magnetic structure in LaFeAsO reported that spatial directions of Fe magnetic moments in the single-stripe-type AFM phase are collinear with the AFM wave vector in the Fe plane, i.e., the orthorhombic \textit{a}-axis, rather than other directions.\cite{18}
This measurement suggests a significant role of the spin-orbit interaction (SOI) in LaFeAsO because the directions of the magnetic moments are coupled to the lattice structure via SOI only.
Effects of SOI on the Fermi surface, the spin-density wave, and the superconducting states were studied theoretically
with symmetry analyses and tight-binding models.\cite{Cvetkovic}
The interplay between SOI and the nematic order was studied theoretically to distinguish their features in the 
paramagnetic-phase electronic band structures.\cite{Fernandes}
The effects of SOI on the electronic band structures were measured recently in the paramagnetic phase of iron-based superconductors by angle-resolved photoemission spectroscopy and compared with DFT 
calculations.\cite{borisenko}

In this paper, we present SOI effects on the electronic and magnetic properties in antiferromagnetic LaFeAsO calculated by first-principles density-functional methods. We show that spatial directions of Fe magnetic moments in the ground state of the single-stripe-type AFM phase are along the in-plane AFM direction, i.e., the orthorhombic $a$ axis, which is consistent with neutron diffraction experiments.\cite{18}
We also show that the energy cost  for aligning Fe magnetic moments perpendicular to the orthorhombic $a$ axis is about 0.1~meV/Fe, indicating that the Fe magnetic moments prefer the orthorhombic $a$ axis direction significantly at temperatures below 1~K.
In addition, we show that SOI splits degenerate bands at a high-symmetry point, Y, in the orthorhombic Brillouin zone by about 60~meV, depending on the spatial direction of the Fe magnetic moment.

In our calculations, we used the generalized gradient approximation\cite{19} to the density functional theory and the projector augmented-wave method\cite{20,21} as implemented in VASP.\cite{22,23}
We regarded $6s^2 5p^6 5d^1$ electrons in La, $3s^2 3p^6 4s^1 3d^7$ in Fe, $4s^2 4p^3$ in As, and $2s^2 2p^4$ in O as valence electrons.
 All of our calculations were performed using a $8\times 8\times 4$ Monkhorst-Pack $k$-point mesh\cite{24} centered at $\Gamma$ for orthorhombic (\textit{Cmme}) structure.
Electronic wavefunctions were expanded with plane waves up to a kinetic-energy cutoff of 400~eV. The tetrahedron method with the Bl\"{o}chl corrections was used for the Brillouin zone integration.\cite{25} With SOI in the total-energy functional, we carefully checked the convergence of the self-consistent calculation of the electron density in order to distinguish small difference in the total energy caused by difference in the spatial direction of the Fe magnetic moment.

\begin{figure} % FIG. 1
\begin{center}
\epsfig{file=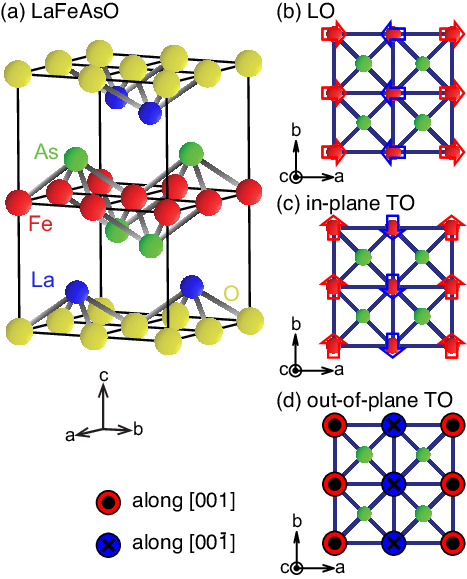,width=8.5cm,angle=0,clip=} % for double column format
\end{center}
\caption{Atomic and magnetic structures in LaFeAsO.
(a) An orthorhombic unit cell of LaFeAsO in the single-stripe-type AFM phase. The unit cell contains four formula units, i.e., 16 atoms. (b) Longitudinal ordering (LO), (c) in-plane 
transverse ordering (TO), and (d) out-of-plane TO of Fe magnetic moments in the FeAs layer.
In LO, in-plane TO, and out-of-plane TO, spatial directions of Fe magnetic moments are parallel or anti-parallel with the orthorhombic $a$, $b$, and $c$ directions, respectively.
In (b) and (c), directions of Fe magnetic moments are denoted by arrows.
In (d), red and blue dots represent Fe magnetic moments along the $+c$ and $-c$ directions, respectively.}
\end{figure}

We optimized atomic positions in LaFeAsO using a 32-atom orthorhombic supercell, $a\times  b\times  2c$, which is twice the structure shown in Fig.~1(a) along the $c$ axis.
During the optimization, the supercell lattice constants were fixed to $a = 5.7063$ \AA, $b = 5.6788$ \AA, and $c = 8.7094$ \AA, which are the measured values at 2~K,\cite{18} and 
the single-stripe-type AFM order was imposed, where Fe magnetic moments is antiferromagnetic along the $a$ direction and ferromagnetic along the $b$ direction.
Atomic positions were fully optimized until residual forces on them were less than 0.001~eV/\AA.
SOI was considered during the structural optimization, but obtained equilibrium positions of atoms depend negligibly on spatial directions of Fe magnetic moments as long as they are in the single-stripe-type AFM phase.

\begin{table}
\caption{Calculated total energies, $E_\mathrm{tot}$, and Fe magnetic moments $m_\mathrm{Fe}$, in LaFeAsO with various spin-configurations, obtained including SOI. Considered spatial directions of Fe magnetic moments are LO, in-plane TO, and out-of-plane TO, as shown in Fig.~1. 
For each spin-configuration, we consider AFM and FM orderings along the $c$ axis; in the former Fe moment directions are in phase along the $c$ axis, while in the latter they are out of phase along the $c$ axis.
The total energy of LO with the AFM ordering along the $c$ axis is set to zero.
Without SOI, all spin-configurations have the same total energy and the same Fe magnetic moment, 1.449~$\mu_\mathrm{B}$/Fe.}
\setlength{\tabcolsep}{3mm}
\renewcommand{\arraystretch}{1.2}
\begin{tabular}{c c c c}
\hline\hline
\multicolumn{2}{c}{spin} & $E_\mathrm{tot}-E_\text{LO}$ & $m_\mathrm{Fe}$ \\
\multicolumn{2}{c}{configuration} & (meV/Fe) & ($\mu_\mathrm{B}$/Fe) \\
\hline
\multirow{3}{10mm}{AFM \\along\\ $c$ axis} & LO & 0.00 & 1.439 \\
%\cline{2-4}
 & in-plane TO & 0.13 & 1.442 \\
%\cline{2-4}
 & out-of-plane TO & 0.12 & 1.439 \\
\hline
\multirow{3}{10mm}{FM \\along\\ $c$ axis} & LO & 0.00 & 1.439 \\
%\cline{2-4}
 & in-plane TO & 0.13 & 1.442 \\
%\cline{2-4}
 & out-of-plane TO & 0.12 & 1.439 \\
\hline\hline
\end{tabular}
\end{table}

To investigate effects of SOI on the Fe magnetic-moment directions, we consider three different configurations, as shown in Figs.~1(b)-(d), where the Fe magnetic moments are along the orthorhombic $a$, $b$, and $c$ directions, respectively.
In Fig.~1(b), the Fe magnetic moments are in a longitudinal ordering (LO) in the sense that they are parallel or anti-parallel to the in-plane AFM ordering vector, i.e., the $a$ axis in our supercell.
In Figs.~1(c) and (d), the Fe magnetic moments are in in-plane and out-of-plane transverse orderings (TOs), respectively, in the sense that they are perpendicular to the in-plane AFM ordering vector and they are either in the Fe plane or out of the plane.
Although the spatial directions of Fe magnetic moments are different in the three configurations, all of them have the same single-stripe-type AFM phase in the sense that neighboring Fe magnetic moments are in opposite directions along the $a$ direction while they are in the same direction along the $b$ direction.
For each of the three configurations, we consider two different inter-Fe-plane magnetic orderings: ferromagnetic (FM) and AFM orderings between Fe magnetic moments neighboring along the $c$ direction.

In our density-functional calculations, we did not impose any constraint on the spatial directions of the Fe magnetic moments; however, whenever we took one of the three configurations as initial guess for the spin density at the begining of the self-consistent iteration in our calculation, the spatial directions of the Fe magnetic moments were stationary during the iteration until the self-consistency was reached. In addition, everytime when the initial directions of the Fe magnetic moments were slightly away from the LO case, the directions of the Fe magnetic moments changed gradually during the self-consistent iteration, converging to the LO case. The LO case is the lowest-energy configuration, as discussed below.

Table~I shows our calculational results for the total energy and Fe magnet moments of LaFeAsO including SOI. The LO case is the lowest-energy configuration while the in-plane TO and the out-of-plane TO are higher in energy than the LO case by 0.13~meV and 0.12~meV per Fe atom, respectively. The total energies in our calculations are independent of the $c$-axis ordering of Fe magnetic moments because of rather large distance between Fe planes in LaFeAsO.

\begin{figure}% FIG. 2
\centering
\epsfig{file=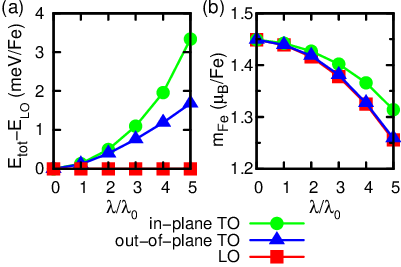,width=8.5cm,angle=0,clip=true}  % for double column format
\caption{
SOI-strength dependences of the total-energy difference and the Fe magnetic moment for three different spin-configurations in LaFeAsO.
(a) The total energy difference ($E_\mathrm{tot}-E_\mathrm{LO}$)
 and (b) the Fe magnetic moment ($m_\mathrm{Fe}$)
are plotted as functions of the SOI strength, $\lambda$. Here, $\lambda_0$ is the original SOI strength.
Red, green, and blue lines represent results of LO, in-plane TO, and out-of-plane TO, respectively. In (a), the total energy of LO ($E_\mathrm{LO}$)
is used as the reference for each $\lambda$ value.}
\end{figure}

Table~I also shows calculated Fe magnetic moments, which are 1.439~$\mu_\mathrm{B}$/Fe for LO, 1.442~$\mu_\mathrm{B}$/Fe for the in-plane TO, and 1.439~$\mu_\mathrm{B}$/Fe for the out-of-plane TO. These values are not affected by whether the Fe magnetic moments are ordered along the $c$ axis ferromagnetically or antiferromagnetically. These magnetic moments are quite greater than the measured value, as discussed below; yet, the dependence of the total energy on the magnetic-moment direction agrees with experiment. Meanwhile, without SOI, the Fe magnetic moment is 1.449~$\mu_\mathrm{B}$/Fe regardless of the moment direction.

Our result that LO is the lowest-energy configuration of Fe magnetic moments in LaFeAsO agrees with the neutron diffraction experiments,\cite{18} where spatial directions of Fe magnetic moments at 2~K in LaFeAsO were claimed to be along the AFM direction in the Fe plane, which corresponds to the $a$ direction in our notation. 
Since our calculated energy gain of LO is about 0.1~meV/Fe and it corresponds to a temperature of about 1~K, our result implies that LO is thermodynamically preferred at temperatures below 1~K.

The neutron diffraction experiments also reported that the size of the Fe magnetic moment in LaFeAsO is
0.36~$\mu_\mathrm{B}$/Fe at 8~K (Ref.~\onlinecite{6}) and
 0.63~$\mu_\mathrm{B}$/Fe at 2~K (Ref.~\onlinecite{18}). Thus
our calculated sizes of Fe magnetic moments are at least about 0.8~$\mu_B$ larger than the measured values, 
but it is consistent with previous DFT results.\cite{7,11,cao,14,15,ma,yang}
This well-known discrepancy between measurements and DFT results suggests that real samples of
LaFaAsO have strong fluctuations in the magnetic moments or correlation effects, which is beyond the description of the standard DFT.\cite{cricchio,yin}

\begin{figure} % FIG. 3
\centering
\epsfig{file=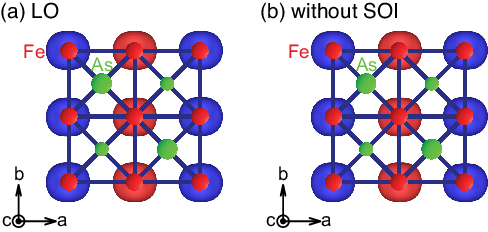,width=8.5cm,angle=0,clip=} % for double column format
\caption{Calculated spin densities in LaFeAsO in the single-stripe-type AFM phase, represented with iso-surfaces. 
(a) The $a$-axis component of the spin density in the LO case obtained with SOI. (b) The spin density obtained without SOI, drawn for comparison.
In (a) and (b), small red and green dots represent Fe and As atoms, respectively, and
 large  blue and red sphere-like
surfaces represent positive and negative values of the spin
density, respectively.}
\end{figure}

\begin{figure}% FIG. 4
\centering
\epsfig{file=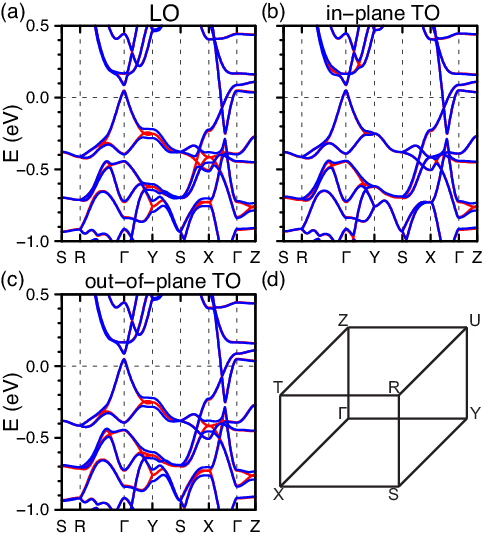,width=8.5cm,angle=0,clip=true}  % for double column format
\caption{Electronic energy bands along the high symmetry lines in LaFeAsO for three different spin-configurations: (a) LO, (b) in-plane TO, and (c) out-of-plane TO.
In (a)-(c), blue and red lines represent band dispersions obtained with and without SOI, respectively. 
(d) One eighth of the first Brillouin zone, where high symmetry points and lines are marked. 
In (a) and (c), splitting of degenerate bands at the Y point by SOI is about 60 meV.
In (d), $\Gamma$X, $\Gamma$Y, and $\Gamma$Z lines are along the orthorhombic $a$, $b$, and $c$ axes in Fig.~1, respectively.}
\end{figure}

To check the validity of the small total-energy difference of about 1~meV/Fe of the three configurations, we re-calculated the total energies as we increased the SOI strength by multiplying a constant factor to the SOI term in the Hamiltonian. We obtained that
as the SOI strength increases, the total-energy differences increase, with the LO case still remaining as the lowest-energy configuration, as shown in Fig.~2(a).
This validates our result that the LO case is the lowest-energy configuration.
We also notice that SOI reduces Fe magnetic moments, as shown in Fig.~2(b).

As shown in Table~I, the magnitudes of Fe magnetic moments depend very weakly on the spatial directions of the moments, and the absence and presence of SOI produces only 0.01~$\mu_\mathrm{B}$/Fe difference.
For further comparison, we obtained and compared the spatial distributions of the spin densities in LaFeAsO with and without SOI. As shown in Fig.~3,
the spin densities with and without SOI are quite similar to each other.

Now we consider effects of SOI on the electronic band structures in LaFeAsO.
Figure~4 shows the band structures in orthorhombic LaFeAsO, where blue and red lines show the band structures with and without SOI, respectively.
Since most parts of the band structures without SOI coincide with those with SOI and the former is drawn prior to the latter, the band structures without SOI, plotted in red, are visible only at $k$-points where SOI produces significant difference.
While band structures without SOI are consistent with previous theoretical results,\cite{7,16,yang} SOI splits some
degenerate bands, as shown in Fig.~4.

Figure~5 shows symmetry operations in LaFeAsO structure. The atomic
structure of LaFeAsO is invariant under 
two mirror operations M$_x^\text{As}$ and M$_y^\text{As}$, and four
rotation operations R$_x^\text{Fe}$, R$_y^\text{Fe}$,
R$_z^\text{Fe}$, and R$_z^\text{As}$. Here, M$_x^\text{As}$ 
(M$_y^\text{As}$) is the mirror
operation with respect to a plane containing As atoms and normal to the $x$ ($y$) axis, R$_x^\text{Fe}$ (R$_y^\text{Fe}$, R$_z^\text{Fe}$) is the 180-degree rotation with respect to an axis through Fe atoms in parallel with the $x$ ($y$, $z$) axis,
and R$_z^\text{As}$ is the 180-degree rotation with respect to an
axis through As atoms in parallel with the $z$ axis.

\begin{figure}% FIG. 5
\centering
\epsfig{file=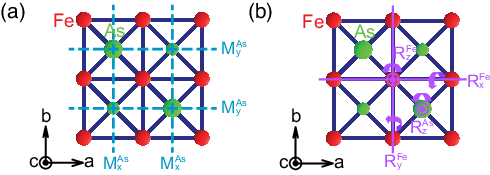,width=8.5cm,angle=0,clip=true}  % for double column format
\caption{
Schematic diagrams of symmetries in nonmagnetic LaFeAsO: (a) mirror and (b) rotation symmetries.
In (a), M$_x^\text{As}$ and M$_y^\text{As}$ are mirror planes at an As atom in parallel with the $bc$ and $ac$ planes, respectively.
In (b), R$_x^\text{Fe}$, R$_y^\text{Fe}$, and R$_z^\text{Fe}$ are 180$^\circ$ rotation with respect to axes at an Fe atom in parallel with the $a$, $b$, and $c$ axes, respectively. R$_z^\text{As}$ is 180$^\circ$ rotation with respect to an axis at an As atom in parallel with the $c$ axis. Here, the $x$, $y$, and $z$ axes are along the $a$, $b$, and $c$ axes, respectively.}
\end{figure}

 When LaFeAsO is antiferromagnetic, it has lower symmetry.
In the LO case where magnetic moments are along the $x$ direction ($\vec{M} \parallel \hat{x}$), Hamiltonian
is invariant under R$_x^\text{Fe}$ and R$_z^\text{As}$. 
In the in-plane TO case ($\vec{M} \parallel \hat{y}$), 
Hamiltonian is invariant
under M$_x^\text{As}$, M$_y^\text{As}$, R$_y^\text{Fe}$, and R$_z^\text{As}$. 
In the out-of-plane TO case ($\vec{M} \parallel \hat{z}$), Hamiltonian is
invariant under M$_x^\text{As}$ and R$_z^\text{Fe}$. These are summarized
in Table II.

\begin{table}
\caption{Symmetry operations in the AFM phases of LaFeAsO.
The LO, in-plane TO, and out-of-plane TO are shown in Fig.~1.
The symmetry operations M$_x^\text{As}$, M$_y^\text{As}$,
R$_x^\text{Fe}$, R$_y^\text{Fe}$, R$_z^\text{Fe}$, and R$_z^\text{As}$
are described in Fig.~5.
}
\setlength{\tabcolsep}{2mm}
\renewcommand{\arraystretch}{1.2}
\begin{tabular}{c c c c c c c}
\hline\hline
 & M$_x^\text{As}$ & M$_y^\text{As}$ &
R$_x^\text{Fe}$ & R$_y^\text{Fe}$ &
R$_z^\text{Fe}$ & R$_z^\text{As}$ \\
\hline
LO & no & no & yes & no & no & yes \\
 in-plane TO &  yes & yes & no & yes & no & yes \\
 out-of-plane TO &  yes & no & no & no & yes & no \\
\hline\hline
\end{tabular}
\end{table}

As shown in Fig.~4, SOI splits the degenerate bands at the Y point by about 60 meV in the LO and out-of-plane TO cases except for
the in-plane TO case.
This lift of the degeneracy results from lowering of the symmetry of the system.
In the in-plane TO case [Fig.~4(b)], all bands are at least four-fold degenerate (including the Kramers degeneracy) in  the YURS plane of the Brillouin zone boundary where $k_y=\pm\pi/b$ [Fig.~4(d)]. This is due to the presence of the mirror symmetry M$_y^\text{As}$. 
In the LO and out-of-plane TO cases, where $M_y^\text{As}$ does not exist, the degeneracy in the SY line is lifted, but bands in the RS line are still degenerate by the presence of R$_z^\text{As}$ in the LO case and R$_z^\text{Fe}$ in the out-of-plane TO case, as shown in Figs.~4(a) and (c),
respectively.

In summary, we have investigated the effects of SOI in the magnetic and electronic structures of LaFeAsO using first-principles density functional calculations. 
In our results, the lowest-energy magnetic structure is the longitudinal ordering of Fe magnetic moments where the moments are parallel or anti-parallel with the in-plane AFM ordering direction. 
This is in good agreement with the experimental results\cite{18} although our magnetic moments are larger than the experimental value. 
The longitudinal alignment of Fe magnetic moments has an energy
gain of about 0.1~meV/Fe, equivalent to a temperature of 1~K.
Band structures with SOI show splitting of some degenerate bands at the high-symmetry points in the Brillouin zone by about 60~meV, depending on the spatial direction of the Fe magnetic moments.
We analyzed the effects of SOI on the electronic band structures by considering the symmetry of the AFM phase.

This work was supported by National Research Foundation of Korea (Grant 
No. 2011-0018306). Computational resources have been provided by KISTI 
Supercomputing Center (Project No. KSC-2008-S02-0004).


\begin{thebibliography}{99}

\bibitem{1}
Y. Kamihara, H. Hiramatsu, M. Hirano, R. Kawamura, H. Yanagi, T. Kamiya, 
and H. Hosono, J. Am. Chem. Soc. {\bf 128}, 10012 (2006).

\bibitem{2}
Y. Kamihara, T. Watanabe, M. Hirano, and H. Hosono, 
J. Am. Chem. Soc. {\bf 130}, 3296 (2008).

\bibitem{3}
H. Takahashi, K. Igawa, K. Arii, Y. Kamihara, M. Hirano, and H. Hosono, 
Nature (London) {\bf 453}, 376 (2008).



\bibitem{4}
Z.-A. Ren, W. Lu, J. Yang, W. Yi, X.-L. Shen, Z.-C. Li, G.-C. Che, 
X.-L. Dong, L.-L. Sun, F. Zhou, and Z.-X. Zhao, 
Chin. Phys. Lett. {\bf 25}, 2215 (2008).

\bibitem{5}
I. I. Mazin, D. J. Singh, M. D. Johannes, and M. H. Du, 
Phys. Rev. Lett. {\bf 101}, 057003 (2008). 
% 18 March 2008

\bibitem{6}
C. de la Cruz, Q. Huang, J. W. Lynn, J. Li, W. R. II, J. L. Zarestky, 
H. A. Mook, G. F. Chen, J. L. Luo, N. L. Wang, and P. Dai, 
Nature (London) {\bf 453}, 899 (2008).
% 1 April 2008

\bibitem{yildirim} T. Yildirim, Phys. Rev. Lett. {\bf 101}, 057010 (2008).
% 17 April 2008

\bibitem{7}
Z. P. Yin, S. Leb\'{e}gue, M. J. Han, B. P. Neal, S. Y. Savrasov, 
and W. E. Pickett, Phys. Rev. Lett. {\bf 101}, 047001 (2008).
% 21 April 2008

\bibitem{8}
H.-H. Klauss, H. Luetkens, R. Klingeler, C. Hess, F. J. Litterst, 
M. Kraken, M. M. Korshunov, I. Eremin, S. -L. Drechsler, R. Khasanov, 
A. Amato, J. Hamann-Borrero, N. Leps, A. Kondrat, G. Behr, J. Werner, 
and B. Buchner, Phys. Rev. Lett. {\bf 101}, 077005 (2008).
% 2 May 2008

\bibitem{9}
J. Dong, H. J. Zhang, G. Xu, Z. Li, G. Li,W. Z. Hu, D. Wu, G. F. Chen, 
X. Dai, J. L. Luo, Z. Fang, and N. L. Wang, 
Europhys. Lett. {\bf 83}, 27006 (2008).

\bibitem{10}
B. Lorenz, K. Sasmal, R. P. Chaudhury, X. H. Chen, R. H. Liu, T. Wu, 
and C. W. Chu, Phys. Rev. B {\bf 78}, 012505 (2008).
% 29 May 2008

\bibitem{11}
I. I. Mazin, M. D. Johannes, L. Boeri, K. Koepernik, and D. J. Singh, 
Phys. Rev. B {\bf 78}, 085104 (2008).
% 5 June 2008

\bibitem{12}
W. Z. Hu, J. Dong, G. Li, Z. Li, P. Zheng, G. F. Chen, J. L. Luo, 
and N. L. Wang, Phys. Rev. Lett. {\bf 101}, 257005 (2008).

\bibitem{13}
V. Cvetkovic and Z. Tesanovic, Europhys. Lett. {\bf 85}, 37002 (2009).


\bibitem{Review_P}
J. Paglione and R. L. Greene,
Nat. Phys. {\bf 6}, 645 (2010).

\bibitem{Review_O}
H. Oh, J. Moon, D. Shin, C.-Y. Moon, and H. J. Choi,
Progress in Superconductivity {\bf 13}, 65 (2011) [arxiv:1201.0237].

\bibitem{Review_D}
P. Dai,
Rev. Mod. Phys. {\bf 87}, 855 (2015).

\bibitem{Review_B}
E. Bascones, B. Valenzuela, and M. J. Calder\'on,
C. R. Physique {\bf 17}, 36 (2016).

\bibitem{cao}
C. Cao, P. J. Hirschfeld, and H.-P. Cheng,
Phys. Rev. B {\bf 77}, 220506 (2008).

\bibitem{14}
C.-Y. Moon, S. Y. Park, and H. J. Choi, Phys. Rev. B {\bf 78}, 212507 (2008).

\bibitem{15}
C.-Y. Moon, S. Y. Park, and H. J. Choi, Phys. Rev. B {\bf 80}, 054522 (2009).

\bibitem{16}
F. Ma, W. Ji, J. Hu, Z.-Y. Lu, and T. Xiang, Phys. Rev. Lett. {\bf 102}, 177003 (2009).

\bibitem{17}
C.-Y. Moon and H. J. Choi, Phys. Rev. Lett. {\bf 104}, 057003 (2010).

\bibitem{18}
N. Qureshi, Y. Drees, J. Werner, S. Wurmehl, C. Hess, R. Klingeler, B. B\"{u}chner, 
M. T. Fern\'{a}ndez-D\'{i}az, and M. Braden, Phys. Rev. B {\bf 82}, 184521 (2010).

\bibitem{Cvetkovic}
V. Cvetkovic and O. Vafek, Phys. Rev. B {\bf 88}, 134510 (2013).
% Space group symmetry, spin-orbit coupling, and the low-energy effective Hamiltonian for iron-based superconductors

\bibitem{Fernandes}
R. M. Fernandes and O. Vafek,
%Distinguishing spin-orbit coupling and nematic order in the electronic spectrum of iron-based superconductors
Phys. Rev. B {\bf 90}, 214514 (2014).

\bibitem{borisenko}
S. V. Borisenko, D. V. Evtushinsky, Z.-H. Liu, I. Morozov, R. Kappenberger, S.Wurmehl,
B. B\"uchner, A. N. Yaresko, T. K. Kim, M. Hoesch, T. Wolf, 
and N. D. Zhigadlo, Nat. Phys. {\bf 12}, 311 (2016).

\bibitem{19}
J. P. Perdew, K. Burke, and Y. Wang, Phys. Rev. B {\bf 54}, 16533 (1996).

\bibitem{20}
P. E. Bl\"{o}chl, Phys. Rev. B {\bf 50}, 17953 (1994). 

\bibitem{21}
G. Kresse and D. Joubert, Phys. Rev. B {\bf 59}, 1758 (1999). 

\bibitem{22}
G. Kresse and J. Hafner, Phys. Rev. B {\bf 47}, 558(R) (1993).

\bibitem{23}
G. Kresse and J. Furthm\"{u}ller, Phys. Rev. B {\bf 54}, 11169 (1996).

\bibitem{24}
H. J. Monkhorst and J. D. Pack, Phys. Rev. B {\bf 13}, 5188 (1976).

\bibitem{25}
P. E. Bl\"{o}chl, O. Jepsen, and O. K. Andersen, Phys. Rev. B {\bf 49}, 16223 (1994).



\bibitem{ma}
F. Ma and Z.-Y. Lu,
Phys. Rev. B {\bf 78}, 033111 (2008).

\bibitem{yang}
Y. Yang and X. Hu, J. Appl. Phys. {\bf 106}, 073910 (2009).

\bibitem{cricchio}
F. Cricchio, O. Gr{\aa}n\"as, and L. Nordstr\"om,
Phys. Rev. B {\bf 81}, 140403 (2010).

\bibitem{yin}
Z. P. Yin, K. Haule, and G. Kotliar, Nat. Mater. {\bf 10}, 932 (2011).


\end{thebibliography}
\end{document}